# Adaptive Non-Uniform Compressive Sensing using SOT-MRAM Multibit Crossbar Arrays

Soheil Salehi, *Student Member, IEEE*, and Ronald F. DeMara, *Senior Member, IEEE*

*Abstract*—A Compressive Sensing (CS) approach is applied to utilize intrinsic computation capabilities of Spin-Orbit Torque Magnetic Random Access Memory (SOT-MRAM) devices for IoT applications wherein lifetime energy, device area, and manufacturing costs are highly-constrained while the sensing environment varies rapidly. In this manuscript, we propose the *Adaptive Compressed-sampling via Multibit Crossbar Array (ACMCA)* approach to intelligently generate the CS measurement matrix using a multibit SOT-MRAM crossbar array. SPICE circuit and MATLAB algorithm simulation results indicate that ACMCA reduces reconstruction Time-Averaged Normalized Mean Squared Error (TNMSE) by 5dB on average while providing up to 160μm² area reduction compared to a similar previous design presented in the literature while incurring a negligible increase in the energy consumption of generating the CS measurement matrix.

## I. INTRODUCTION

In recent years, one of the main focuses of research in the Internet of Things (IoT) applications has been optimizing energy consumption while maximizing signal sampling performance and reconstruction accuracy [1]–[4]. Recently, to decrease the energy consumption as well as storage needs and data transmission overheads, Compressive Sensing (CS) approaches are being investigated. Unlike conventional sampling methods that require the sampling to be performed at the Nyquist rate, CS algorithms aim to sample spectrally-sparse wide-band signals close to their information rate. Utilizing CS approaches help mitigate the overhead cost of sampling hardware [5]–[8]. Non-uniform CS algorithms utilize Random Number Generator (RNG) for random sampling of the signal [6]. However, traditional True RNGs (TRNGs) and Pseudo-RNGs (PRNGs) may face challenges such as slow speed, limited quality of randomness, and area and energy consumption overheads incurred due to post-processing requirements [1], [9]. Thus, there is an increased demand for RNG circuits that are energy- and area-efficient and can provide adaptive behavior. Traditional implementation of non-uniform CS algorithms in hardware have been implemented using Complementary Metal Oxide Semiconductor (CMOS) technology [10], [11] and often result in inefficiencies in terms of area and power dissipation. Furthermore, recent advances in spintronics have enabled researchers to design TRNGs using Magnetic Tunnel Junctions (MTJs) [9], stochastic switching in MTJs using sub-threshold voltages [1], [2] precessional switching in MTJs [12], and Voltage-Controlled Magnetic Anisotropy (VCMA) MTJs [13]. However, all of these designs result in area footprint and energy consumption overheads due to their relatively complex hardware.

Herein, we devise a novel circuit-algorithm solution called *Adaptive Compressed-sampling via Multibit Crossbar Array (ACMCA)*, which utilizes non-uniform compressive sensing algorithms along with spin-based hardware circuit to maximize sampling and reconstruction performance and minimize energy consumption and area overheads. The proposed ACMCA approach utilizes Spin Orbit Torque Magnetic Random Access Memory (SOT-MRAM) based multibit resistive devices to generate and store the CS measurement matrix elements. SOT-MRAM-based multibit resistive devices maintain a small area footprint benefits and offer significant reduction in energy consumption [14]. Moreover, SOT-MRAM-based Multibit Cells (SMCs) are utilized in a crossbar array fashion to perform Vector Matrix Multiplications (VMMs) required for sampling and reconstruction of the IoT signals.

The remainder of this paper is organized as follows. In Section II, background and related work are discussed and a detailed description of the SOT-MRAM-based multibit resistive devices utilized herein is provided. The proposed ACMCA approach is described in detail in Section III. Additionally, simulation results and comparisons are presented in Section IV. Finally, Section V concludes this manuscript.

## II. BACKGROUND AND RELATED WORK

### A. Fundamentals of Compressive Sensing

Compressive Sensing (CS) techniques are designed to perform reconstruction algorithms to recover a $k$-sparse signal of length $N$ using $M$ measurements, with $M \ll N$. Based on the definition, a $k$-sparse signal has $k$ non-zero entries in a given basis. Furthermore, the sparsity rate of the signal is defined as $(\frac{k}{N})$. Fundamentally, in order to sample the sparse signal vector, $x \in R^N$, we can use the measurement matrix, $\Phi \in R^{M \times N}$, and the relation $y = \Phi x$ to find the compressed measurements vector, $y \in R^M$. According to the literature, the sparse signal vector, $x \in R^N$, can be recovered from $M$ measurements by solving the basis pursuit problem [15]:

$$\hat{x} = argmin \ \|x\|_1 \ s.t. \ y = \Phi x \quad (1)$$

where $\|x\|_1 = \sum_i |x|$. It has been shown that $\hat{x}$ reconstructs the original signal vector if $\Phi$ satisfies a special condition known as the Restricted Isometry Property (RIP). An $M \times N$ matrix $\Phi$ satisfies RIP($p$) if for any $k$-sparse vector $x$:

$$\|x\|_p(1-\delta) < \|\Phi x\|_p \leq \|x\|_p(1+\delta), 0 < \delta < 1 \quad (2)$$

Furthermore, the sparsity of the signal may be non-uniform and parts of the signal may carry more weight in the reconstruction accuracy, which are called Regions of Interest (RoI) [8], [16]. Thus, it is crucial to employ an adaptive measurement matrix that is non-uniform and allows maximizing performance of reconstruction of the RoI parts of the signal that maintain higher sparsity rates. This can be done by sampling the RoI with higher frequency by adjusting the measurement matrix. Authors in [8] and [16] have verified that RIP condition is satisfied by non-uniform measurement matrices. Thus, non-uniform measurement matrices may be used for sparse signal sampling and reconstruction. Typically, non-uniform CS measurement matrices utilize Bernoulli and Gaussian distributions as shown in Fig. 1. Spectrally sparse signals are utilized in many applications such as frequency hopping communications, musical audio signals, cognitive radio networks, and radar/sonar imaging systems [3].

As mentioned earlier, maximizing non-uniform CS reconstruction accuracy can be done through utilization of a measurement matrix that can adaptively change based on sparse input signal characteristics observed over time [8], [16]. Recent achievements in high-performance sparse signal recovery algorithms utilizing adaptive measurement matrices have shown promising performance improvements [17]–[24]. However, they lack a feasible pathway to implement the proposed algorithm within a hardware fabric considering the signal and hardware constraints or they require extensive hardware support to implement CS techniques. Additionally, previous CS hardware implementations incur significant overheads in terms of area footprint and energy consumption due to use of a large number of CMOS transistors [10], [11]. Thus, in order to reduce area and energy consumption overheads, we devise a low-complexity hardware design.

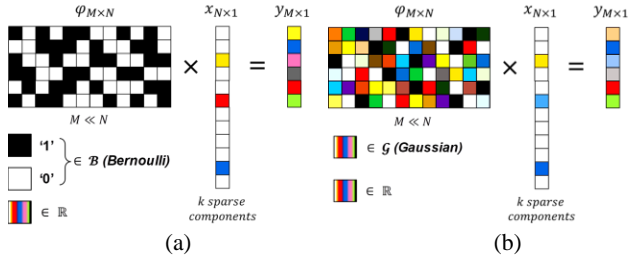

Fig. 1: Compressive Sensing with a (a) Bernoulli measurement matrix and (b) Gaussian non-uniform measurement matrix.

*B. SOT-MRAM-based Multibit Resistive Device*

The SOT-MRAM-based multibit resistive device provides a separate read and write path which will reduce the read error rate, since write operation is performed using the spin Hall Effect (SHE) write mechanism [14]. Additionally, SOT-MRAM is expected to perform better in terms of endurance, power consumption, and speed [25]. The SMCs utilize the intrinsic property that is probabilistic in nature. Authors in [14] have fabricated and characterized an array of nanomagnets with Perpendicular Magnetic Anisotropy (PMA) located on a tantalum layer that acts as a SHE channel. Probabilistic switching of the individual nanomagnets is used to change the state of each SMC. The total magnetization state of the nanomagnets in an SMC can be gradually increased or decreased by applying the proper current pulse through the tantalum layer. It is worth noting that the increase and decrease of the magnetization states of the SMCs is non-linear due to their stochastic nature. However, this can be addressed by modifying the amplitude and duration of the current pulse to update the state of the SMCs. SOT switching of a fabricated Hall bar with a single nanomagnet is performed using current pulses of 50μs width in the presence of an in-plane external field of 20mT in the current direction. Note that the presence of an external magnetic field has been shown to not be a requirement for SOT driven switching of PMA nanomagnets. Some of the SOT switching approaches used structures like wedges [26] or novel GSHE materials such as antiferromagnet PtMn [27] to eliminate the need for utilizing an external magnetic field. It has been demonstrated that a current pulse with width of 50μs can be applied to deterministically reset the state of all nanomagnets to -1. Moreover, a current with varying amplitudes is applied to probabilistically set the state of nanomagnets to +1. Note that higher current amplitude during the set operation will result in higher switching probability. Additionally, reducing the current pulse to 20μs will be required to increase the current amplitude. We can utilize $2^n$ nanomagnets operating in the probabilistic switching regime and working in parallel or series to design an n-bit SMC.

### III. PROPOSED APPROACH

Herein, we propose a novel MRAM-based Adaptive Non-uniform CS approach that utilizes the aforementioned multibit SOT-MRAM resistive devices. We utilize the SOT-MRAM based multibit crossbar array structure as shown in Fig. 2(a) to implement a CS algorithm. The SMCs shown in Fig. 2(b) are used within the crossbar array architecture to generate and store the measurement matrix, which is consisted of three main steps: reset, write, and read operations. In order to write into SMCs, first the write operation control signals, *WWL* and $\overline{WWL}$, enable the write transmission gates, *TGW1* and *TGW2*, which will connect the write path from bit line (*BL*) to source line (*SL*). Then, a write current will be applied with controlled pulses to enable probabilistic switching of the SMCs. To read the value stored in SMCs, the read operation control signals, *RWL* and $\overline{RWL}$, enable the read path from input port (*In*) to output port (*Out*), through the read transmission gates, *TGR1* and *TGR2*. After the read process, the measurement matrix elements can be modified via resetting the state of SMCs. This is performed similar to the write process except with a higher write current amplitude for deterministic switching.

According to the experimental results demonstrated in [14], the authors apply 20 current pulses with 6.97mA magnitude and 20μs width. In this approach, due to the probabilistic nature of switching of the SMC nanomagnets, depending on number of current pulses applied, magnitude of current, and width of pulses, only some of the nanomagnets may switch. Note that since all the current pulses have the same direction, they can only switch the nanomagnets from -1 to +1 state and once a nanomagnet is switched to +1, it will maintain that state in all subsequent current pulses. Herein, we devise a 4-bit SMC using MTJ devices instead of nanomagnets, where the overall resistance of the SMC is distributed between 1KΩ to 5KΩ.

As shown in Fig. 2, our proposed architecture offers control over the number of measurements and signal elements to provide flexibility to adjust to the signal characteristics such as sparsity rate, noise, etc. In particular, due to the non-volatility, zero leakage power dissipation, small area footprint, and instant-on operation features of the SMCs, the unused SMCs can be turned off while incurring nearly zero overhead in terms of energy consumption and area. Thus, we can modify the number of rows in the measurement matrix to increase the number of measurements in order to account for increased sparsity rate. Moreover, we can adjust the number of columns in the measurement matrix in order to increase accuracy of the signal recovery.

Moreover, the SMC crossbar is utilized to perform the VMMs required for compressive sampling and reconstruction of the IoT signals. In the SMC crossbar array, the number of input ports are equal to the number of signal elements while the number of output ports are equal to the number of measurements. Thus, VMM is performed by applying the input signal elements to the input ports of the SMC crossbar array, similar to the one shown in Fig.2(a) as *In0*, and the result of the analog VMM for each measurement will be provided at the output ports of each row of the measurement matrix, similar to the one shown in Fig.2(a) as *Out0*, where the outputs will be connected to a Winner Takes All (WTA) circuit in the output node. Furthermore, the number of signal elements and the number of measurements can be adaptively adjusted, which will enable the algorithm to adapt to the signal characteristics as well as the hardware constraints. In particular, the algorithm can trade off its reconstruction accuracy versus energy consumption by reducing the number of measurements or increase the number of measurements in case of high signal sparsity rate to maximize the reconstruction performance.

Additionally, we introduce an update variable in the algorithm in [17], [28] to modulate the frequency of updates to the measurement matrix while maintaining control over the energy consumption overhead. This variable, which is called γ, determines the update frequency of the measurement matrix and it is modified in each

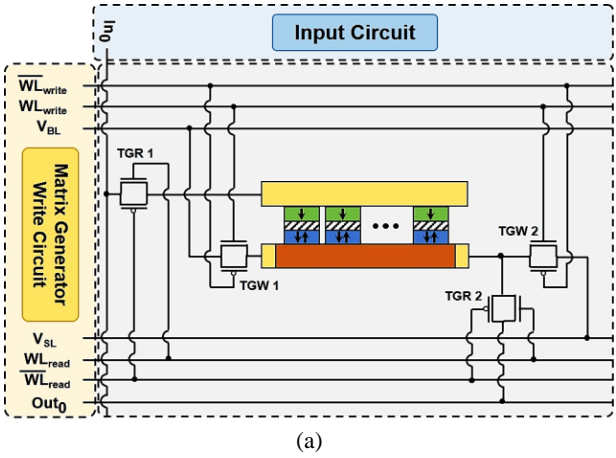

(a)

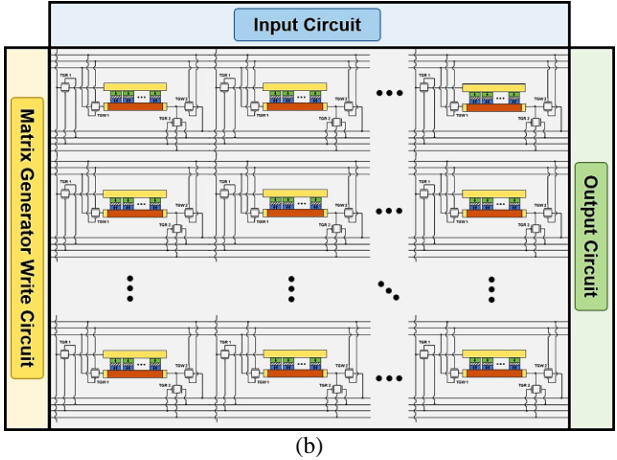

(b)

Fig. 2: Multibit stochastic SOT-MRAM-based (a) single cell, and (b) crossbar array.

Table 1: Comparison of energy needed for VMM in CMOS Crossbar vs. proposed SMC Crossbar.

| Array Size (N×M) | CMOS Crossbar Energy Consumption | SMC Crossbar Energy Consumption | Energy Improvement |
|---|---|---|---|
| 100×25 | 1,177 pJ | 240 pJ | ~5X |
| 200×50 | 4,708 pJ | 968 pJ | ~4.8X |
| 400×100 | 18,832 pJ | 3840 pJ | ~4.9X |

According to our results, energy consumption estimates for VMM using an SMC crossbar array to implement our modified version of the CS algorithm introduced in [17], [28] using a variety of N and M values compared to CMOS crossbar array are listed in Table 1. As shown in Table 1, utilizing the SMC crossbar array for VMM operation provides 5-fold energy consumption reduction compared to a CMOS crossbar array. Furthermore, write energy consumption estimates for populating the measurement matrix is 3nJ for N×M=100×25, 12nJ for N×M=200×50, and 50nJ for N×M=400×100, on average. According to our simulations, total energy consumption estimate for the CS reconstruction algorithm's VMM operation is 7nJ for N×M=100×25, 29nJ for N×M=200×50, and 115nJ for N×M=400×100. The CS algorithms discussed in [17], [28] have been utilized to analyze the performance of ACMCA, as shown in Fig. 3. We performed our simulations using a sparse signal with $N = 400$, sparsity rate of $\frac{k}{N} = 0.1$, various measurement $M = \{40, 60, 80, 100\}$, and RoI occupying 10% of the entire signal. Our Monte Carlo functional simulation results with 100 instances indicate that the proposed approach is capable of decreasing the Time-Averaged Normalized Mean Squared Error (TNMSE) of RoI coefficients up to 1dB for various measurement values using a Bernoulli measurement matrix. Additionally, this offers up to 6dB reduction in TNMSE utilizing Gaussian measurement matrix. Furthermore, this approach eliminates the need for an external iteration according to the energy budget. In particular, if γ is greater than the energy budget, we will update the measurement matrix in every iteration of the algorithm. Moreover, if γ is less than the energy budget and greater than the critical energy, we will reduce the frequency of updates to the measurement matrix. Additionally, if γ is less than the critical energy, we will further decrease the frequency of updates made to the measurement matrix. However, doing so might increase the overall reconstruction error rate. It is worth noting that the increase in the overall reconstruction error rate can be considered negligible since the reconstruction error rate of the RoI will still be reduced compared to the uniform CS, which is the goal of the algorithm. Moreover, maximizing the update frequency of the measurement matrix will result in minimizing reconstruction error rate at the cost of increased energy consumption. This adaptive behavior will enable the designer to account for hardware constraints. Multi-objective optimization of energy parameters could be performed to improve performance while saving energy [29]. Furthermore, approaches such as [30] can be utilized to address issues affecting the read reliability via dynamic resource allocations in the crossbar array.

## IV. SIMULATION RESULTS

We have performed SPICE circuit and MATLAB algorithm simulations to evaluate the behavior and efficiency of our proposed ACMCA approach. We have used the Pin-Sim framework [31] along with the SMC device simulation, demonstration, and experimental results provided in [14] in our simulations to validate the proposed ACMCA approach and extract area and energy consumption estimates.

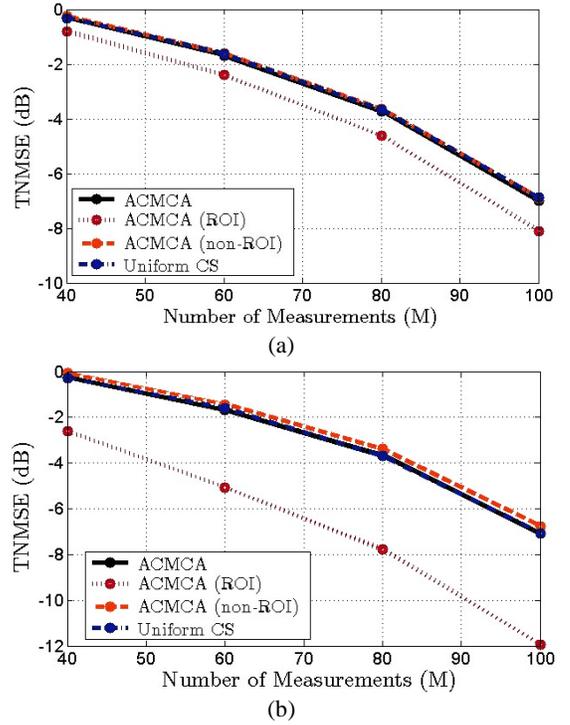

Fig. 3: TNMSE vs. Number of Measurements (M), for a signal with N=400, $\frac{k}{N} = 0.1$, and RoI occupying 10% using (a) Bernoulli measurement matrix, and (b) Gaussian measurement matrix.

TRNG circuit to generate random numbers to populate the measurement matrix.

Proposed approach offers reduced reconstruction TNMSE by 5dB on average while providing up to 160μm$^2$ area reduction compared to the similar previous design presented in [15] according to estimate based on the transistor count in 14nm technology node. ACMCA using N×M=400×100 increases energy consumption by 7-fold on average to populate the measurement matrix. However, the measurement matrix is populated infrequently and once an estimate of the RoI is achieved, there will only be small updates to a few SMCs and not the entire matrix. Thus, the energy consumption overhead of populating the measurement matrix can be considered negligible. Additionally, using the intrinsic probabilistic behavior of SMCs, ACMCA eliminates need for TRNG circuits, which are usually bulky and consume a significant amount of energy when implemented in CMOS.

V. CONCLUSION

We have devised a spin-based non-uniform compressive sensing circuit-algorithm solution that considers the signal dependent constraint as well as hardware limitations called *Adaptive Compressed-sampling via Multibit Crossbar Array (ACMCA)*. High payoff considerations to leverage for device hardware optimization which are advanced herein include the signal sparsity and noise levels. Our circuit-algorithm simulation results indicate that the proposed approach offers reduced reconstruction TNMSE by 6dB on average compared to uniform CS approaches. Additionally, due to infrequent update of the measurement matrix, the energy consumption overhead of populating the measurement matrix can be considered negligible. Moreover, ACMCA eliminates need for TRNG circuit via utilizing the intrinsic probabilistic behavior of SMCs.

ACKNOWLEDGMENT

This work was supported in part by the National Science Foundation (NSF) through ECCS 1810256.

REFERENCES

[1] Y. Qu, J. Han, B. F. Cockburn, W. Pedrycz, Y. Zhang, and W. Zhao, "A True Random Number Generator Based on Parallel STT-MTJs," in *Proceedings of the Conference on Design, Automation & Test in Europe (DATE '17)*, 2017, pp. 606–609.
[2] Y. Wang, H. Cai, L. Alves De Barros Naviner, J.-O. Klein, and W. Zhao, "A Novel Circuit Design of True Random Number Generator Using Magnetic Tunnel Junction," in *IEEE/ACM International Symposium on Nanoscale Architectures (NANOARCH)*, 2016, pp. 123–128.
[3] S. Salehi, M. B. Mashhadi, A. Zaeemzadeh, N. Rahnavard, and R. F. DeMara, "Energy-Aware Adaptive Rate and Resolution Sampling of Spectrally Sparse Signals Leveraging VCMA-MTJ Devices," *IEEE J. Emerg. Sel. Top. Circuits Syst.*, vol. 8, no. 4, pp. 679–692, Dec. 2018.
[4] S. Salehi, R. Zand, A. Zaeemzadeh, N. Rahnavard, and R. F. DeMara, "AQuRate: MRAM-based Stochastic Oscillator for Adaptive Quantization Rate Sampling of Sparse Signals," in *in Proceedings of the 2019 on Great Lakes Symposium on VLSI*, 2019, pp. 359–362.
[5] S. Sarvotham, D. Baron, R. G. Baraniuk, S. Sarvotham, D. Baron, and R. G. Baraniuk, "Measurements vs. Bits: Compressed Sensing meets Information Theory," *Allert. Conf. Commun. Control Comput.*, Sep. 2006.
[6] N. Rahnavard, A. Talari, and B. Shahrasbi, "Non-uniform compressive sensing," in *Communication, Control, and Computing (Allerton), 2011 49th Annual Allerton Conference on*, 2011, pp. 212–219.
[7] A. Zaeemzadeh, M. Joneidi, and N. Rahnavard, "Robust Target Localization Based on Squared Range Iterative Reweighted Least Squares," in *2017 IEEE 14th International Conference on Mobile Adhoc and Sensor Systems*, 2017.
[8] B. Shahrasbi and N. Rahnavard, "Model-Based Nonuniform Compressive Sampling and Recovery of Natural Images Utilizing a Wavelet-Domain Universal Hidden Markov Model," *IEEE Trans. Signal Process.*, vol. 65, no. 1, pp. 95–104, Jan. 2017.
[9] D. Vodenicarevic *et al.*, "Low-Energy Truly Random Number Generation with Superparamagnetic Tunnel Junctions for Unconventional Computing," *Phys. Rev. Appl.*, vol. 8, no. 5, p. 054045, Nov. 2017.
[10] D. Bellasi, L. Bettini, T. Burger, Q. Huang, C. Benkeser, and C. Studer, "A 1.9 GS/s 4-bit sub-Nyquist flash ADC for 3.8 GHz compressive spectrum sensing in 28 nm CMOS," in *2014 IEEE 57th International Midwest Symposium on Circuits and Systems (MWSCAS)*, 2014, pp. 101–104.
[11] T.-F. Wu, C.-R. Ho, and M. S.-W. Chen, "A Flash-Based Non-Uniform Sampling ADC With Hybrid Quantization Enabling Digital Anti-Aliasing Filter," *IEEE J. Solid-State Circuits*, vol. 52, no. 9, pp. 2335–2349, 2017.
[12] N. Rangarajan, A. Parthasarathy, and S. Rakheja, "A Spin-based True Random Number Generator Exploiting the Stochastic Precessional Switching of Nanomagnets," *J. Appl. Phys.*, vol. 121, no. 22, p. 223905, Jun. 2017.
[13] H. Lee, F. Ebrahimi, P. K. Amiri, and K. L. Wang, "Design of high-throughput and low-power true random number generator utilizing perpendicularly magnetized voltage-controlled magnetic tunnel junction," *AIP Adv.*, vol. 7, no. 5, p. 055934, May 2017.
[14] V. Ostwal, R. Zand, R. DeMara, and J. Appenzeller, "A novel compound synapse using probabilistic spin-orbit-torque switching for MTJ based deep neural networks," *IEEE J. Explor. Solid-State Comput. Devices Circuits (JxCDC), in-press*, Nov. 2019.
[15] S. Salehi, A. Zaeemzadeh, A. Tatulian, N. Rahnavard, and R. F. DeMara, "MRAM-Based Stochastic Oscillators for Adaptive Non-Uniform Sampling of Sparse Signals in IoT Applications," in *2019 IEEE Computer Society Annual Symposium on VLSI (ISVLSI)*, 2019, pp. 403–408.
[16] A. Zaeemzadeh, M. Joneidi, and N. Rahnavard, "Adaptive non-uniform compressive sampling for time-varying signals," in *2017 51st Annual Conference on Information Sciences and Systems (CISS)*, 2017, pp. 1–6.
[17] A. Zaeemzadeh, J. Haddock, N. Rahnavard, and D. Needell, "A Bayesian Approach for Asynchronous Parallel Sparse Recovery," in *52nd Asilomar Conference on Signals, Systems, and Computers*, 2018, pp. 1980–1984.
[18] A. Gilbert and P. Indyk, "Sparse recovery using sparse matrices," in *Proceedings of the IEEE*, 2010, vol. 98, no. 6, pp. 937–947.
[19] S. Jafarpour, W. Xu, B. Hassibi, and R. Calderbank, "Efficient and robust compressed sensing using optimized expander graphs," *IEEE Trans. Inf. theory*, vol. 55, no. 9, pp. 4299–4308, 2009.
[20] H. T. Kung and S. J. Tarsa, "Partitioned compressive sensing with neighbor-weighted decoding," in *MILITARY COMMUNICATIONS CONFERENCE, 2011 - MILCOM 2011*, 2011, pp. 149–156.
[21] L. Gan, "Block Compressed Sensing of Natural Images," in *15th International Conference on Digital Signal Processing*, 2007, pp. 403–406.
[22] Y. Yu, B. Wang, and L. Zhang, "Saliency-based compressive sampling for image signals," *IEEE Signal Process. Lett.*, vol. 17, no. 11, pp. 973–976, 2010.
[23] Y. Shen, W. Hu, R. Rana, and C. T. Chou, "Nonuniform Compressive Sensing for Heterogeneous Wireless Sensor Networks," in *IEEE Sensors Journal,* 2013, vol. 13, no. 6, pp. 2120–2128.
[24] Y. Liu, X. Zhu, L. Zhang, and S. H. Cho, "Expanding Window Compressed Sensing for Non-Uniform Compressible Signals," *Sensors*, vol. 12, no. 10, pp. 13034–13057, 2012.
[25] M. Cubukcu *et al.*, "SOT-MRAM 300mm integration for low power and ultrafast embedded memories K.," *IEEE Trans. Magn.*, vol. 54, no. 4, pp. 81–82, 2018.
[26] Y. W. Oh *et al.*, "Field-free switching of perpendicular magnetization through spin-orbit torque in antiferromagnet/ferromagnet/oxide structures," *Nat. Nanotechnol.*, vol. 11, no. 10, pp. 878–884, 2016.
[27] S. Fukami, C. Zhang, S. Duttagupta, A. Kurenkov, and H. Ohno, "Magnetization switching by spin-orbit torque in an antiferromagnet-ferromagnet bilayer system," *Nat. Mater.*, vol. 15, no. 5, pp. 535–541, 2016.
[28] A. Zaeemzadeh, M. Joneidi, N. Rahnavard, and G. J. Qi, "Co-SpOT: Cooperative Spectrum Opportunity Detection Using Bayesian Clustering in Spectrum-Heterogeneous Cognitive Radio Networks," *IEEE Trans. Cogn. Commun. Netw.*, vol. 4, no. 2, pp. 206–219, Jun. 2018.
[29] R. S. Oreifej, C. A. Sharma, and R. F. DeMara, "Expediting GA-based evolution using group testing techniques for reconfigurable hardware," in *Proceedings of the 2006 IEEE International Conference on Reconfigurable Computing and FPGAs, ReConFig*, 2006, pp. 106–113.
[30] K. Zhang, G. Bedette, and R. F. Demara, "Triple Modular Redundancy with Standby (TMRSB) supporting dynamic resource reconfiguration," in *AUTOTESTCON (Proceedings)*, 2007, pp. 690–696.
[31] R. Zand, K. Y. Camsari, S. Datta, and R. F. Demara, "Composable probabilistic inference networks using MRAM-based stochastic neurons," *ACM J. Emerg. Technol. Comput. Syst.*, vol. 15, no. 2, Mar. 2019.